\documentclass[journal]{IEEEtran}
\usepackage{helvet}
\usepackage{courier}
\usepackage{amsmath}
\usepackage{amssymb}
\usepackage{graphicx}
\usepackage{subfigure}
\usepackage{algorithm}
\usepackage{algorithmic}
\usepackage{multirow}
\usepackage{makecell} 
\usepackage{diagbox} 
\usepackage{cases}
\usepackage{threeparttable}
\usepackage{caption}

\newtheorem{myPro}{Problem}
\newtheorem{myDef}{Definition}

\begin{document}
%
\title{TraLFM: Latent Factor Modeling of Traffic Trajectory Data}
%
%
%

\author{Meng~Chen, Xiaohui~Yu,~\IEEEmembership{Member,~IEEE,}
        and~Yang~Liu,~\IEEEmembership{Member,~IEEE}
\thanks{M. Chen is with the School of Software, Shandong University, 250002 Jinan, Shandong, China (e-mail: mchen@sdu.edu.cn).}
\thanks{X. H. Yu is with the School of Information Technology, York University, M3J 1P3 Toronto, ON, Canada (e-mail: xhyu@yorku.ca).}
\thanks{Y. Liu is with the Department of Physics and Computer Science, Wilfrid Laurier University, N2L 3C5 Waterloo, ON, Canada (email: yliu@sdu.edu.cn).}
}

\maketitle

\begin{abstract}
The widespread use of positioning devices (e.g., GPS) has given rise to a vast body of human movement data, often in the form of trajectories. Understanding human mobility patterns could benefit many location-based applications. In this paper, we propose a novel generative model called \textbf{TraLFM} via latent factor modeling to mine human mobility patterns underlying traffic trajectories. TraLFM is based on three key observations: (1) human mobility patterns are reflected by the sequences of locations in the trajectories; (2) human mobility patterns vary with people; and (3) human mobility patterns tend to be cyclical and change over time. Thus, TraLFM models the joint action of sequential, personal and temporal factors in a unified way, and brings a new perspective to many applications such as latent factor analysis and next location prediction. We perform thorough empirical studies on two real datasets, and the experimental results confirm that TraLFM outperforms the state-of-the-art methods significantly in these applications.
\end{abstract}


\section{Introduction}\label{intro}

\begin{figure}
\centering
\includegraphics[width=0.4\textwidth]{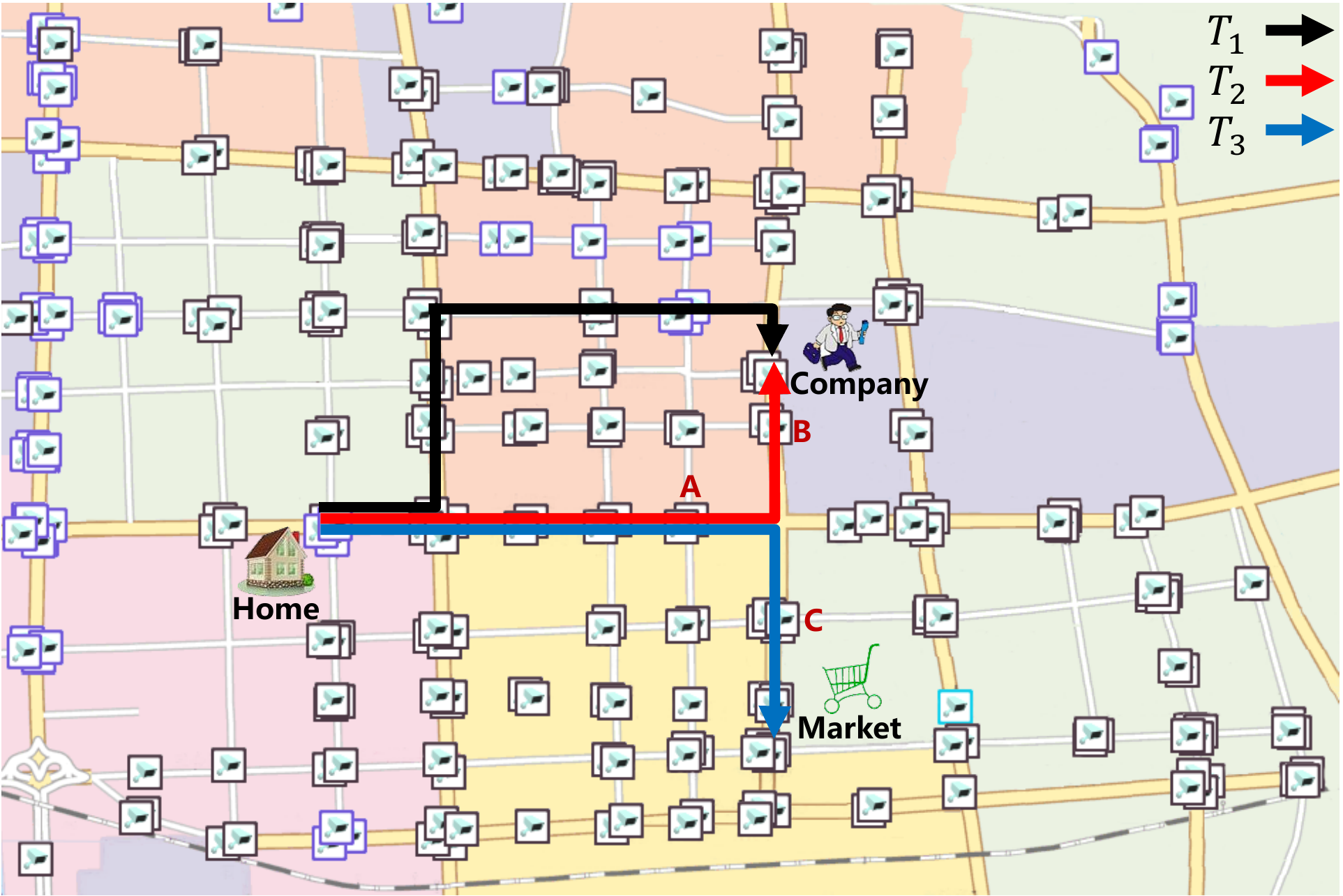}
\caption{A real traffic surveillance system. There may be four cameras in one intersection, monitoring the vehicles from different directions. Each camera represents a location. There are three trajectories $T_1$, $T_2$, $T_3$, which are represented by the black line, the red line, and the blue line, respectively.}
\label{fig:introduction}
\end{figure}

The increasing prevalence of video capturing equipments and electronic dispatch systems has made it possible to collect a deluge of traffic trajectory data. For example, as shown in Fig.~\ref{fig:introduction}, with the deployment of traffic surveillance cameras on roads, vehicles are photographed when they pass the cameras and structured vehicle passage records (VPRs) are subsequently extracted from the pictures using optical character recognition (OCR). As another example, with the widespread adoption of electronic dispatch systems, these mobile data terminals are installed in each taxi and typically provide information on GPS (Global Positioning System) localization and taximeter state. Such types of data can be considered as consisting of records with at least three attributes: \textit{object ID}, \textit{location ID}, and \textit{time-stamp}, and a consecutive sequence of such records from the same object constitute a trajectory. Learning from such trajectory data is an important task, and substantial progress in this domain can have a strong impact on many applications ranging from urban computing, location-based recommendations and trajectory prediction \cite{chen2018pcnn,qiao2015traplan,yin2017spatial,bao2017planning,chen2018mpe,zhao2016predicting}.

In this paper, we are particularly interested in modeling these latent factors that drive people to move from one location to another with the traffic trajectory data. In general, latent factors cannot be explicitly spelled out, and we may consider them as the hidden semantic structures underlying trajectories and the summarization of the semantic information of locations, similar to the topics of documents in the area of text mining. For example, as shown in Fig.~\ref{fig:introduction}, one may drive to his/her workplace every workday morning with the ``working'' activity, and go to a market from home with the ``shopping'' activity on weekends. Furthermore, users have their own driving habits, e.g., some people prefer highway and they may choose $T_1$ to go to the company, and others may choose $T_2$. Such ``activities'' and ``habits'' can be considered as part of the latent factors driving people's movement.

These latent factors could help us understand human mobility patterns, and are highly valuable to emerging location-based services, such as location prediction and user profiling. For instance, supposed a user has arrived at location $A$ (see Fig.~\ref{fig:introduction}), the most probable next location will be location $B$ with the ``working'' activity, and location $C$ with the ``shopping'' activity. Further, each user has its own mobility patterns, which could be reflected by these latent factors, and we could make user clustering correspondingly. Consequently, learning these latent factors could help make location prediction accurately and enhance the quality of user profiling.


However, the problem of modeling human mobility patterns from traffic trajectory data is very challenging, due to the following important factors:
\begin{itemize}
\item \textbf{Sequential patterns.} Human mobility patterns are reflected by the whole trajectory, rather than any individual location. For example, the trajectory from home to company in Fig.~\ref{fig:introduction} is a routine driving route of a user every workday morning, and with this we know that the latent factor may be working. If we only focus on the individual locations (e.g., $A$, $B$, or a series of intermediate locations in the trajectory), we cannot understand human mobility patterns.
\item  \textbf{Personal tendencies.} Human mobility patterns vary from one person to another. For instance, the movements of employees are mainly relevant to working on workdays, however, the retirees are more likely to go shopping or take exercise.
\item \textbf{Temporal influence.} People tend to exhibit nonuniform and periodic moving behaviors at different time, e.g., a person may leave home either in early Monday mornings for work or in late weekend nights for entertainment.
\item \textbf{Missing semantics.} The semantics of locations are often missing in the traffic trajectory data. For example, the locations of the surveillance cameras could be anywhere on the roads, and it is very difficult to assign exact semantic information to each location.
\end{itemize}

To tackle these challenges, we follow a generative approach to learn human mobility patterns via latent factor modeling. A similar generative approach has been successfully used in text mining for inferring the hidden structure underlying texts, with the most prominent being the topic models \cite{blei2003latent,wang2006topics,girolami2005sequential,rosen2004author}. In such models, typically each document is associated with multiple topics, and each word in a document is generated by sampling first from the topic distributions and then from the word distributions of the chosen topic. However, modeling the latent factors of traffic trajectories presents the following brand new challenges that cannot be tackled by simply extending/modifying those models developed for text mining.

First, in topic models, words are usually assumed to be unordered in a document; however, locations in a trajectory are temporally-ordered and the order matters in modeling human mobility patterns. For example, the trajectory for a user going to work from home may contain exactly the same set of locations as the trajectory for her/him to go home after work, but the sequences of locations contained therein are directly opposite. Second, human mobility patterns are affected by both personal and temporal factors, and failing to consider them may hinder the performance. Finally, a trajectory contains three attributes: object, location and time-stamp, and we need to capture the interactions among the three attributes simultaneously.

Therefore, we propose a novel generative model called \textbf{TraLFM} for mining human mobility patterns via latent factor modeling. Intuitively, we assume that trajectories are generated in the following way: a user first samples the latent factors; then, depending on the chosen ones, she samples the appropriate trajectory. In particular, the latent factors are related to the following three features: (1) the object (personal information), (2) the location sequence (sequential information), and (3) the time-stamp attached to every sequence (temporal information). Each latent factor has a different emitting distribution over location sequences, objects, and time; the observed trajectory is considered to be generated by a mixture of latent factors.

We employ Gibbs sampling to perform parameter estimation, and conduct thorough experimental studies on two real datasets: taxi trajectory data which consists of the complete trips of 442 taxis running in the city of Porto (Portugal) for a complete year and VPR data generated by over 35,000 vehicles from a traffic surveillance system installed in a metropolitan area in China. We show that TraLFM is able to obtain better latent factor coherence underlying the mobility behaviors, and we demonstrate their effectiveness on the task of next location prediction. The experimental results confirm the superiority of our proposals over alternative methods.

To summarize, we make the following contributions to mining and modeling human mobility patterns:
\begin{itemize}
\item We propose a model named TraLFM to better understand human mobility patterns via latent factor modeling of traffic trajectory data. With the help of TraLFM, we can understand the mobility patterns and predict next locations.
\item We incorporate personal and temporal information into the generative process of trajectories, such that we are not only able to capture the meaningful mobility patterns, but also able to determine which objects (time periods) demonstrate a more similar moving behavior.
\item We conduct extensive experiments with two large-scale and real-world datasets. The results verify the effectiveness of TraLFM and show remarkable improvement compared with the baselines.

\end{itemize}
\textbf{Roadmap.} The rest of our paper is organized as follows. We survey the related work in Section 2. Section 3 gives the preliminaries. We introduce our TraLFM in Section 4 and discuss parameter estimation in Section 5. The experimental results and performance analysis are presented in Section 6. Section 7 concludes this paper.

\section{Related Work}

In this section, we first review the studies on mobility pattern mining, and then discuss the recent progress of generative methods and the related applications.

\subsection{Mobility Pattern Mining}\label{relatedwork}

There have appeared a considerable body of works that aim at modeling individuals' mobility behaviors with traffic trajectory data (e.g., taxi trajectories). For example, Yuan et al. \cite{yuan2015discovering} present a collaborative-filtering-based approach to learn the location and mobility semantics from human trajectories including POIs, road networks, public transit data and taxi trajectories, and identify functional zones (e.g., educational areas, entertainment areas) in a city. Chen et al. \cite{chen2015mining} consider both individual and collective movement patterns and propose a next location predictor based on Markov models. Qiao et al. \cite{qiao2018predicting} design a prefix-projection-based trajectory prediction algorithm called PrefixTP based on frequent sequential pattern mining to predict long-term trajectories. In our study, we are concerned with mining human mobility patterns with the support of latent factors from the traffic trajectory data, which has not been studied by previous work.

In addition, there also have been some studies \cite{kim2015toptrac,yin2016joint,yuan2015and,zhang2016gmove} that aim to understand individuals' mobility behaviors using spatio-temporal data in location-based social networks, such as Twitter, Weibo, and Foursquare. For example, Kim et al. \cite{kim2015toptrac} propose a novel probabilistic model to discover patterns in the trajectories of geo-tagged text messages. Yuan et al. \cite{yuan2015and} present a probabilistic model $W^{4}$ to exploit text messages associated with geographic information, posting time and user ids to discover human mobility behaviors. Zhang et al. \cite{zhang2016gmove} propose a group-level mobility modeling method to model group-level human mobility using geo-tagged social media. However, this type of strategies cannot be trivially applied on traffic trajectory data. First, check-in data focus on the individual POI with semantic category information, which represents the activity (e.g., shopping) of people in that location, whereas traffic data focus on the visiting order of locations in a trajectory, and the mobility patterns of a trajectory is not the integration of semantic information of a series of locations. Second, each POI has detailed semantic information (e.g., Mexican restaurant) or multi-level category information, which is missing in the traffic trajectory data.

\subsection{Generative Models}

Our work is closely related to topic mining in the area of text mining, as the task of latent factor modeling in movements naturally corresponds to that of finding topics in text. Both probabilistic latent semantic analysis(PLSA) \cite{hofmann1999probabilistic} and latent dirichlet allocation (LDA) \cite{blei2003latent} have been popular methods for exploratory analysis of text. There also appear a series of topic models which consider author's information \cite{rosen2004author}, temporal information \cite{wang2006topics}, and word order \cite{wallach2006topic}. Rosen-Zvi et al. \cite{rosen2004author} propose the Author-Topic model (ATM) which extends LDA to include authorship information. In ATM, each author is associated with a multinomial distribution over topics and each topic is associated with a multinomial distribution over words. Wang and McCallum \cite{wang2006topics} present Topics over Time (TOT), a topic model which models time jointly with word co-occurrence patterns. Wallach \cite{wallach2006topic} introduces a Bigram Topic Model that extends LDA by incorporating a notion of word order. Compared with the models \cite{blei2003latent,rosen2004author,wang2006topics,wallach2006topic}, our model considers more aspects: 1) we take the order into consideration, which plays a pivotal role in modeling human mobility patterns; 2) we consider the user's information and the temporal information jointly.

There also exist a few studies that use generative models to mine the trajectories without semantic information. For instance, Long et al. \cite{long2012exploring} use LDA model to discover the local geographic topics from the check-in data. Specifically, they consider the venue in a check-in record as a word and a user's trajectory as a document. Farrahi et al. \cite{farrahi2011discovering} propose to incorporate location sequences into the LDA and Author-Topic model to discover location-driven routines. Shen et al. \cite{shen2009topic} use global transition probability to model the Markov dependence of data, and propose a probabilistic mixture model T-BiLDA to mine sequences of temporal activities. In addition, Yin et al. \cite{yin2016joint, wang2017st} propose multiple generative models to simultaneously mine the semantic, temporal and spatial patterns of users' check-in activities to make point-of-interest (POI) recommendation. However, these methods use some factors (e.g., semantic information) that do not exist in traffic trajectory data. So we simplify the spatial-temporal sparse additive generative model (ST-SAGE) \cite{wang2017st} to fit our data, which takes into account both user interests and temporal dynamics of user behaviors. We will compare our proposed model with these aforementioned methods in the experiments.

\section{Preliminaries}
For the sake of convenience, we define the following terms used throughout the paper and list the notations in Tab.~\ref{tab:notation}.
\begin{table}
\centering
\caption{Notations and descriptions.}
\renewcommand{\arraystretch}{1.2}
\begin{tabular}{ l| l}
\Xhline{1pt}
Notations & Descriptions \\
\hline
$M, K$ &  the total number of trajectories and latent factors  \\
$T_{s}$ &  the number of $r$th-order sequences in a trajectory\\
$T, u, z$  &  trajectory, trajectory unit, latent factor\\
$o, l, t, s$ &  object, location, time, $r$th-order sequence\\
$\vec{s}, \vec{o}, \vec{t}, \vec{z}$ &  vector for $r$th-order sequence, object, time, latent factor\\
\hline
$\vec{\theta}$ &  the trajectory specific latent factor distribution\\
$\vec{\phi}$ &  the latent factor specific sequence distribution\\
$\vec{\psi}$ &  the latent factor specific object distribution\\
$\vec{\varphi}$ &  the latent factor specific time distribution\\
$\vec{\alpha}, \vec{\beta}, \vec{\eta}, \vec{\gamma}$ &  Dirichlet priors \\
\Xhline{1pt}
\end{tabular}
\label{tab:notation}
\end{table}	

\begin{myDef} For an object $o$, its \textbf{trajectory} $T$ is defined as a time-ordered sequence of location-time pairs: $\langle(l_{1}, t_{1}),(l_{2}, t_{2}), \ldots, (l_{n}, t_{n})\rangle$, where $l_i$ and $t_i$ are location and time-stamp respectively ($1\le i\le n$).
\end{myDef}

For a trajectory, we require that the time difference between two neighboring locations be less than a threshold. The value of the threshold is usually data-dependent. Here we set the threshold at one hour according to the suggestion in \cite{lou2013traffic}.

\begin{myDef} For a trajectory $\langle(l_{1}, t_{1}),(l_{2}, t_{2}), \ldots, (l_{n}, t_{n})\rangle$, a \textbf{$r$th-order sequence} $s_i$ is defined as a length-$(r+1)$ sequence of locations starting at location $i$: $s_i= \langle l_{i},\ldots, l_{i+r}\rangle$, where $1 \leq i \leq n-r$.
\end{myDef}

Note that the sequence hereinafter is also referred to as $r$th-order sequence when the meaning is clear from the context. We discretize the time of a day into hourly-based bins, and use the bin that the average time-stamp of all the locations in the $r$th-order sequence belongs to as the time of $s_i$, denoted as $\overline{t}_i$. The detailed method of processing time is elaborated in Section 4.3. This way, we turn the trajectory $\langle(l_{1}, t_{1}),(l_{2}, t_{2}), \ldots, (l_{n}, t_{n})\rangle$ into $\langle(s_{1},\overline{t}_{1}), (s_{2},\overline{t}_{2}), \ldots, (s_{n-r}, \overline{t}_{n-r})\rangle$.

\begin{myDef} For a trajectory $\langle(s_{1}, \overline{t}_{1}), (s_{2}, \overline{t}_{2}), \ldots,\\ (s_{n-r}, \overline{t}_{n-r})\rangle$ of object $o$, we define $(s_{i}, o, \overline{t}_{i})$ in the trajectory as a \textbf{trajectory unit} $u$, where $1 \leq i \leq n-r$.
\end{myDef}

\begin{myPro} Given the historical trajectories, \textbf{Latent Factor Modeling} aims at mining human mobility patterns by introducing the latent factors, and learning the trajectory specific latent factor distribution $\vec{\theta}$, the latent factor specific sequence distribution $\vec{\phi}$, the latent factor specific object distribution $\vec{\psi}$, and the latent factor specific time distribution $\vec{\varphi}$.
\end{myPro}

\section{Latent Factor Modeling of Trajectories}
To capture the latent factors underlying the traffic trajectories, we start with modeling them with the $r$th-order sequences. As distributions over the latent factors vary with objects and time periods, to reflect the influence of personal and temporal information, we then refine our model by introducing the latent factor specific object distribution and the latent factor specific time distribution to the basic model respectively. Finally, we develop an integrated model by studying the interactions of location sequences, objects, and time simultaneously.

\subsection{Modeling Sequential Patterns}
Different from topic models considering only the occurrences rather the sequential patterns of objects (e.g., the bag-of-words semantics in text mining) \cite{blei2003latent,wang2006topics}, understanding human mobility patterns might be highly influenced by the order of locations therein. Therefore, we opt to learn the latent factors with $r$th-order sequences instead of single locations in the trajectory.

We start with a basic model. As trajectories can be represented as a mixture of latent factors, we choose a multinomial distribution $\vec{\theta}$ to represent the trajectory specific latent factor distribution, generated from a symmetric Dirichlet ($\vec{\alpha}$) prior. Here we choose Dirichlet as the prior distribution, as it is conjugate to the multinomial distribution and has finite dimensional sufficient statistics. Similarly, we use a multinomial distribution $\vec{\phi}$ to express the latent factor specific sequence distribution, which is sampled from a Dirichlet with parameter $\vec{\beta}$. With this model, a trajectory $T$ can be considered as generated by repeatedly emitting $r$th-order sequences $\vec{s}$. Each such sequence is generated by first sampling from a prior distribution over the possible latent factor $\vec{z}$, and then for the latent factor obtained, drawing a sampled $r$th-order sequence from a multinomial distribution $\vec{\phi}$.

Several assumptions are made here. First, the number of the latent factors ($K$) is assumed known and fixed. Second, the latent factor specific sequence distribution $\vec{\phi}$ is parameterized by a $K\times S$ matrix ($S$ is the total number of $r$th-order sequences), which we treat as a fixed quantity that needs to be estimated. Finally, $T_{s}$, the number of $r$th-order sequences in a trajectory $T$, is an ancillary variable and is independent of other generative variables (e.g., $\vec{z}$, $\vec{\phi}$).

Given the parameters $\vec{\alpha}$ and $\vec{\beta}$, the trajectory specific latent factor distribution $\vec{\theta}$, a set of $K$ latent factors $\vec{z}$, and $T_{s}$ $r$th-order sequences $s_{i}$ in the trajectory $T$, the marginal distribution of trajectory $T$ can be represented as
\begin{equation}
\small
\begin{aligned}
&p(\lbrace s_{i}\rbrace_{i=1}^{T_{s}}|T;\vec{\alpha}, \vec{\beta}) = \\
&\int p(\vec{\theta}|\vec{\alpha})\prod_{i=1}^{T_{s}}\sum_{k=1}^{K} p(z_{k}|T;\vec{\theta}) p(s_{i}|z_{k};\vec{\beta})d\vec{\theta}.
\end{aligned}
\end{equation}

\subsection{Modeling Personal Tendencies}
Intuitively, people have their personal preferences. For example, the employees often drive to work every workday morning, whereas the retirees are more likely to go shopping or take exercise. To this end, we extend the basic model by incorporating additional object information.

Except the trajectory specific latent factor distribution $\vec{\theta}$ and latent factor specific sequence distribution $\vec{\phi}$, we add a latent factor specific object distribution $\vec{\psi}$, which is a multinomial distribution over a group of objects, sampled from a Dirichlet with parameter $\vec{\eta}$. We assumes the following generative process for each trajectory: choose $\vec{\theta} \sim Dir(\vec{\alpha})$ first; then for each of the $T_{s}$ sequences, pick a latent factor from $\vec{\theta}$ and sample a $r$th-order sequence and an object from the selected latent factor. Here, note that the distribution $\vec{\psi}$ is parameterized by a $K\times O$ matrix. Integrating $p(\vec{o}|\vec{z};\vec{\eta})$ with $p(\lbrace s_{i}\rbrace_{i=1}^{T_{s}}|T;\vec{\alpha}, \vec{\beta})$, we obtain the marginal distribution of a trajectory $T$:
\begin{equation}
\small
\begin{aligned}
&p( \lbrace s_{i},o_{i}\rbrace_{i=1}^{T_{s}}|T;\vec{\alpha},\vec{\beta},\vec{\eta}) = \\
&\int p(\vec{\theta}|\vec{\alpha})\prod_{i=1}^{T_{s}}\sum_{k=1}^{K} p(z_{k}|T;\vec{\theta}) p(s_{i}|z_{k};\vec{\beta}) p(o_{i}|z_{k};\vec{\eta})d\vec{\theta}.
\end{aligned}
\end{equation}

\subsection{Modeling Temporal Factor}
Time also plays an important role in understanding human mobility patterns, and such information can be captured with the time-stamp that is associated with each location. For example, a person is more likely to leave home in early Monday mornings for work instead of entertainment. Thus, we incorporate the temporal information into the basic model.

We notice that people's mobility behaviors demonstrate two unique characteristics: (1) the movement patterns of people tend to be cyclical; (2) people's patterns on weekdays are significantly different from those in weekends. Accordingly, we discretize the time of a day into equi-sized bins, and in each of which, we set a multinomial distribution over the latent factors. We further assume all workdays share a similar pattern, and only distinguish time bins in weekdays from weekends. For example, we set \textit{2 hours} as a bin, and thus obtain 24 bins (12 for weekdays and 12 for weekends) in total. We will evaluate the effect of the size of time bins in the experiments.

In this way, we model both location sequences and their time-stamps. We add a latent factor specific time distribution $\vec{\varphi}$, which is a multinomial distribution generated from a symmetric Dirichlet ($\vec{\gamma}$). $\vec{\varphi}$ is parameterized by a fixed $K\times B$ matrix. The marginal distribution of trajectory $T$ can be represented as
\begin{equation}
\small
\begin{aligned}
&p( \lbrace s_{i},t_{i}\rbrace_{i=1}^{T_{s}}|T;\vec{\alpha},\vec{\beta}, \vec{\gamma}) = \\
&\int p(\vec{\theta}|\vec{\alpha})\prod_{i=1}^{T_{s}}\sum_{k=1}^{K} p(z_{k}|T;\vec{\theta}) p(s_{i}|z_{k};\vec{\beta}) p(t_{i}|z_{k};\vec{\gamma})d\vec{\theta}.
\end{aligned}
\end{equation}

\subsection{The Complete Model}
\begin{figure}
\centering
\includegraphics[width=0.35\textwidth]{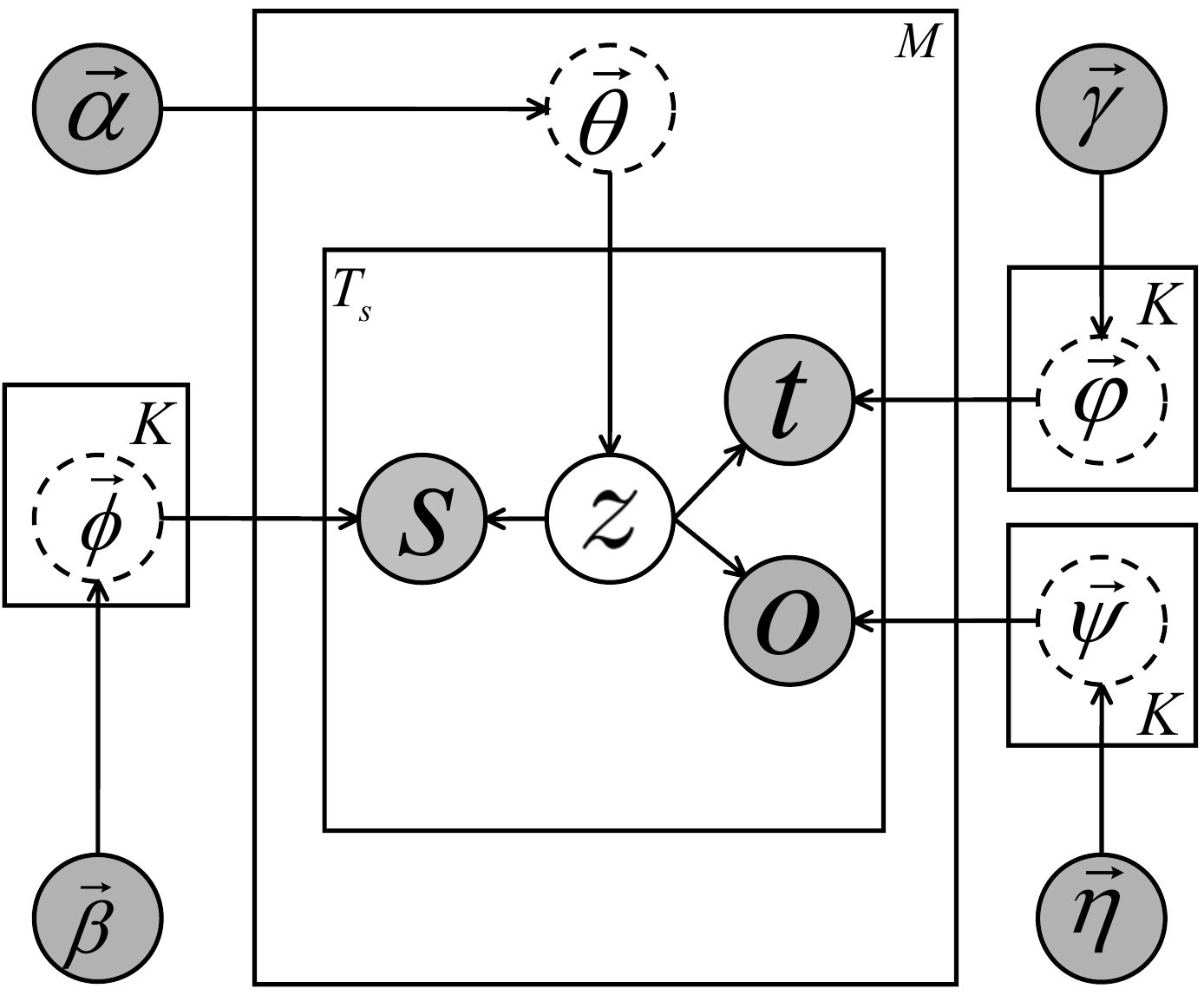}
\caption{Graphical model representation of TraLFM.}
\label{fig:model}
\end{figure}

We now present TraLFM to model the latent factors by jointly modeling object, $r$th-order sequences and time-stamps in a trajectory. In this model, we assume that each trajectory has a set of latent factors $\vec{z}$. For a trajectory $T=m$, the latent factors are generated from a trajectory specific latent factor distribution $\vec{\theta}_{m}$. For each latent factor of $\vec{z}$, objects are generated from the distribution $\vec{\psi}_{z}$, the $r$th-order sequences are generated from the distribution $\vec{\phi}_{z}$, and the time of each sequence is generated from the distribution $\vec{\varphi}_{z}$. The graphical model of this process is shown in Fig.~\ref{fig:model}. As depicted in the figure, there are three levels to the TraLFM representation. The parameters $\vec{\alpha}, \vec{\beta}, \vec{\eta}, \vec{\gamma}$ are prior-level parameters, which are sampled only once in the process of generating a set of trajectories. $\vec{\theta}$ is a trajectory-level variable, sampled once per trajectory. The variables $\vec{z}, \vec{s}, \vec{o}, \vec{t}$ are unit-level variables, sampled once for each trajectory unit in a trajectory.

The generative process of a trajectory can be described as follows:
\begin{itemize}
\item Draw $K$ latent factor specific sequence distributions $\vec{\phi}_{z}$ from a Dirichlet prior $\vec{\beta}$,  $\vec{\phi}_{z} \sim Dir(\vec{\beta})$;
\item Draw $K$ latent factor specific object distributions $\vec{\psi}_{z}$ from a Dirichlet prior $\vec{\eta}$, $\vec{\psi}_{z} \sim Dir(\vec{\eta})$;
\item Draw $K$ latent factor specific time distributions $\vec{\varphi}_{z}$ from a Dirichlet prior $\vec{\gamma}$,  $\vec{\varphi}_{z} \sim Dir(\vec{\gamma})$;
\item For the trajectory $T=m$, draw a trajectory specific latent factor distribution $\vec{\theta}_{m}$ from a Dirichlet prior $\vec{\alpha}$; then for each trajectory unit in the trajectory:
\begin{itemize}
\item[(1)] Draw a latent factor $z_{m}$ from $\vec{\theta}_{m}$;
\item[(2)] Draw a $r$th-order sequence $s_{m}$ from  $\vec{\phi}_{z_{m}}$;
\item[(3)] Draw an object $o_{m}$ from  $\vec{\psi}_{z_{m}}$;
\item[(4)] Draw a time $t_{m}$ from $\vec{\varphi}_{z_{m}}$.
\end{itemize}
\end{itemize}

Given the parameters $\vec{\alpha}, \vec{\beta}, \vec{\eta}, \vec{\gamma}$, the trajectory specific latent factor distribution $\vec{\theta}$, a set of $K$ latent factors $\vec{z}$, and $T_{s}$ trajectory units $(s_{i},o_{i},t_{i})$ in trajectory $T$, the marginal distribution of $T$ in TraLFM can be represented as
\begin{equation}
\small
\begin{aligned}
&p( \lbrace s_{i},o_{i},t_{i}\rbrace_{i=1}^{T_{s}}|T;\vec{\alpha},\vec{\beta}, \vec{\eta}, \vec{\gamma}) = \\
&\int p(\vec{\theta}|\vec{\alpha})\prod_{i=1}^{T_{s}}\sum_{k=1}^{K} p(z_{k}|T;\vec{\theta}) p(s_{i}|z_{k};\vec{\beta}) p(o_{i}|z_{k};\vec{\eta}) p(t_{i}|z_{k};\vec{\gamma}) d\vec{\theta}.
\end{aligned}
\end{equation}

TraLFM contains a total of $\vec{\theta}, \vec{\phi}, \vec{\psi}, \vec{\varphi}$ variables, where $\vec{\theta}$ is treated as a $K$-parameter hidden random variable which is not related to the training set. The distributions $\vec{\phi}$, $\vec{\psi}$, $\vec{\varphi}$ are parameterized by the fixed $K\times S, K\times O, K\times B$ matrix, respectively.  The number of parameters to be estimated is $(K+KS+KO+KB)$. Furthermore, the parameters in TraLFM do not grow with the size of the training set. Note that in our study, on one hand, the number of latent factors is bounded by a pre-determined constant; nonetheless, a non-parametric Bayesian method that automatically determines the number of latent factors can also be applied \cite{song2005modeling}. On the other hand, values of the Dirichlet parameters have been assumed to be known in TraLFM, but they can be learned from the data to increase model quality, and one can refer to \cite{gregor2005parameter} for the detailed hyper parameters estimation.

\section{Learning and Parameter Estimation}
After elaborating the generative process of a trajectory through TraLFM, we devote to the procedures for inference and parameter estimation in this section. The pivotal problem for learning TraLFM is to compute the posterior distribution of the latent factors given a trajectory, i.e., $p\left( \vec{z}|\vec{s},\vec{o},\vec{t} \right)$. Unfortunately, exact inference is intractable for the posterior distribution. The solution to this is to use approximate inference algorithms, such as Gibbs sampling \cite{gregor2005parameter}. As a special case of Markov-chain Monte Carlo (MCMC) simulation, Gibbs sampling is usually used for approximate inference in topic models (e.g., LDA) due to the relatively simple algorithms. To this end, we also choose this method to make a derivation.

To derive a Gibbs sampler, we first identify the variables and parameters in our model. The hidden variable is $\vec{z}$, i.e., the latent factors that appear with the trajectory units; the observed variables are the trajectory units $(\vec{s},\vec{o},\vec{t})$. The parameters need to be estimated are the distributions $\vec{\theta}$, $\vec{\phi}$, $\vec{\psi}$, and $\vec{\varphi}$. The target of inference is the distribution $p(\vec{z}|\vec{s},\vec{o},\vec{t})$:
\begin{equation}
\tiny
\begin{aligned}
p\left( \vec{z}|\vec{s},\vec{o},\vec{t} \right) = &\dfrac{p\left( \vec{z},\vec{s},\vec{o},\vec{t} \right)}{p\left( \vec{s},\vec{o},\vec{t} \right)}\\
= &\dfrac{\prod_{i=1}^{T_{s}}p(z_{i},s_{i},o_{i},t_{i})}{\prod_{i=1}^{T_{s}} \sum_{k=1}^{K} p(z_{i}=k,s_{i},o_{i},t_{i})},
\end{aligned}
\label{equ:ztu}
\end{equation}
which is directly proportional to the joint distribution $p\left( \vec{z},\vec{s},\vec{o},\vec{t} \right)$. Since the variables $\vec{s}$, $\vec{o}$ and $\vec{t}$ are assumed as independent, the joint distribution can be represented as
\begin{equation}
\tiny
p\left( \vec{z},\vec{s},\vec{o},\vec{t} \right) =  p(\vec{s}|\vec{z})
p(\vec{o}|\vec{z})p(\vec{t}|\vec{z})p(\vec{z}).
\label{equ:zsot}
\end{equation}

The first term, $p(\vec{s}|\vec{z})$, can be derived from a multinomial on the observed sequence given the associated latent factor:
\begin{equation}
\tiny
\begin{aligned}
p(\vec{s}|\vec{z},\vec{\phi}) =& \prod_{k=1}^{K} \prod_{i:z_{i}=k} p(s_{i}=s|z_{i}=k)\\
=& \prod_{k=1}^{K} \prod_{s=1}^{S} \vec{\phi}_{k,s}^{n_{k}^{s}},
\label{s|z}
\end{aligned}
\end{equation}
where $n_{k}^{s}$ denotes the number of times that the $r$th-order sequence $s$ has been observed with the latent factor $k$, and $S$ is the total number of $r$th-order sequences. The target distribution $p(\vec{s}|\vec{z},\vec{\beta})$ is obtained by integrating over $\vec{\phi}$, which can be done using Dirichlet integrals with the product over $\vec{z}$.
\begin{equation}
\tiny
\begin{aligned}
p(\vec{s}|\vec{z},\vec{\beta}) =& \int p(\vec{s}|\vec{z},\vec{\phi})p(\vec{\phi}|\vec{\beta})d\vec{\phi}\\
=& \prod_{k=1}^{K} \dfrac{\Delta(\vec{n}_{k}^{s}+\vec{\beta})}{\Delta(\vec{\beta})},
\label{equ:sz}
\end{aligned}
\end{equation}
where $\vec{n}_{k}^{s}={\{n_{k}^{s}\}}_{s=1}^{S}$ and $\Delta(\vec{\beta})$ is the Dirichlet delta function with parameters $\vec{\beta}$.

Analogous to $p(\vec{s}|\vec{z},\vec{\beta})$, the latent factor specific object distribution $p(\vec{o}|\vec{z},\vec{\eta})$, the latent factor specific time distribution $p(\vec{t}|\vec{z},\vec{\gamma})$ and the trajectory specific latent factor distribution $p(\vec{z}|\vec{\alpha})$ can be computed. The joint distribution therefore becomes:
\begin{equation}
\tiny
\begin{aligned}
p\left( \vec{z},\vec{s},\vec{o},\vec{t} \right) &= \prod_{k=1}^{K} \dfrac{\Delta(\vec{n}_{k}^{s}+\vec{\beta})}{\Delta(\vec{\beta})} \cdot \prod_{k=1}^{K} \dfrac{\Delta(\vec{n}_{k}^{o}+\vec{\eta})}{\Delta(\vec{\eta})} \cdot \prod_{k=1}^{K} \dfrac{\Delta(\vec{n}_{k}^{t}+\vec{\gamma})}{\Delta(\vec{\gamma})} \cdot \\
& \quad \prod_{m=1}^{M} \dfrac{\Delta(\vec{n}_{m}+\vec{\alpha})}{\Delta(\vec{\alpha})},
\end{aligned}
\end{equation}
where $\vec{n}_{k}^{o}={\{n_{k}^{o}\}}_{o=1}^{O}$, $\vec{n}_{k}^{t}={\{n_{k}^{t}\}}_{t=1}^{B}$, $\vec{n}_{m}={\{n_{m}^{k}\}}_{k=1}^{K}$, and $M$ is the total number of trajectories.

From the joint distribution, we can compute the conditional distribution $p\left( z_{i}=k|\vec{z}_{-i},\vec{s},\vec{o},\vec{t} \right)$ for a trajectory unit with index $i$. The Gibbs sampler draws the latent factor for the $i$-th trajectory unit with the update equation:
\begin{equation}
\tiny
\begin{aligned}
&p\left( z_{i}=k|\vec{z}_{-i},\vec{s},\vec{o},\vec{t} \right) \\
&=\dfrac{p(\vec{s}|\vec{z})}{p(\vec{s}_{-i}|\vec{z}_{-i})p(s_{i})}  \dfrac{p(\vec{o}|\vec{z})}{p(\vec{o}_{-i}|\vec{z}_{-i})p(o_{i})}  \dfrac{p(\vec{t}|\vec{z})}{p(\vec{t}_{-i}|\vec{z}_{-i})p(t_{i})}  \dfrac{p(\vec{z})}{p(\vec{z}_{-i})} \\
&\propto \dfrac{\Delta(\vec{n}_{k}^{s}+\vec{\beta})}{\Delta(\vec{n}_{k,-i}^{s}+\vec{\beta})} \dfrac{\Delta(\vec{n}_{k}^{o}+\vec{\eta})}{\Delta(\vec{n}_{k,-i}^{o}+\vec{\eta})} \dfrac{\Delta(\vec{n}_{k}^{t}+\vec{\gamma})}{\Delta(\vec{n}_{k,-i}^{t}+\vec{\gamma})} \dfrac{\Delta(\vec{n}_{m}+\vec{\alpha})}{\Delta(\vec{n}_{m,-i}+\vec{\alpha})}\\
&= \dfrac{n_{k,-i}^{s}+\beta_{s}}{\sum_{s=1}^{S} n_{k,-i}^{s}+\beta_{s}} \cdot \dfrac{n_{k,-i}^{o}+\eta_{o}} {\sum_{o=1}^{O}n_{k,-i}^{o}+\eta_{o}}\cdot
 \dfrac{n_{k,-i}^{t}+\gamma_{t}}{\sum_{t=1}^{B}n_{k,-i}^{t}+\gamma_{t}} \cdot \\
& \quad \dfrac{n_{m,-i}^{k}+\alpha_{k}}{\sum_{k=1}^{K}n_{m,-i}^{k}+\alpha_{k}},
\end{aligned}
\label{equ:newtopic}
\end{equation}
where $z_{i}=k$ indicates that the assignment of the $i$-th trajectory unit in the trajectory is the latent factor $k$. $\vec{z}_{-i}$ represents all the latent factors assignments except for the $i$-th trajectory unit, and the same apply to $\vec{s}_{-i}$, $\vec{o}_{-i}$ and $\vec{t}_{-i}$. Furthermore, $n_{m,-i}^{k}$ is the number of times the latent factor $k$ has appeared in the $m$-th trajectory, $n_{k,-i}^{s}$ is the number of sequence $s$ assigned to the latent factor $k$,  $n_{k,-i}^{o}$ is the number of object $o$ assigned to the latent factor $k$, and $n_{k,-i}^{t}$ is the number of  time $t$ assigned to the latent factor $k$. However, all of them do not contain the current instance. $S$, $O$ and $B$ are the number of unique $r$th-order sequences, objects and time bins, respectively.

Finally, we need to obtain the multinomial parameter sets $\vec{\phi}$, $\vec{\psi}$, $\vec{\varphi}$, and $\vec{\theta}$ that correspond to the latent factors, $\vec{z}$. In light of the definitions of multinomial distributions with Dirichlet prior, we apply Bayes' rule on the latent factor $z=k$,
\begin{equation}
\tiny
\begin{aligned}
p(\vec{\phi}_{k}|\vec{z},\vec{s},\vec{\beta}) &= \dfrac{1}{Z_{\phi_{k}}} \prod_{i:z_{i}=k} p(s_{i}|\vec{\phi}_{k})p(\vec{\phi}_{k}|\vec{\beta}) \\
&= Dir(\vec{\phi}_{k}|\vec{n}_{k}^{s}+\vec{\beta})  \\
p(\vec{\theta}_{m}|\vec{z}_{m},\vec{\alpha}) &= \dfrac{1}{Z_{\theta_{m}}} \prod_{n=1}^{T_{s}} p(z_{mn}|\vec{\theta}_{m})p(\vec{\theta}_{m}|\vec{\alpha}) \\
&= Dir(\vec{\theta}_{m}|\vec{n}_{m}+\vec{\alpha}),
\end{aligned}
\end{equation}
where $\vec{n}_{m}$ is the vector of latent factor observation counts for trajectory $m$ and $\vec{n}_{k}^{s}$ that of
sequence observation counts for the latent factor $k$. $p(\vec{\psi}_{k}|\vec{z},\vec{o},\vec{\eta})$ and
$p(\vec{\varphi}_{k}|\vec{z},\vec{t},\vec{\gamma})$ have the similar formulas. Using the expectation of the Dirichlet distribution, on these results yields:

\begin{equation}
\small
\label{equ:para2}
\begin{aligned}
\vec{\theta}_{m,k} &= \dfrac{n_{m}^{k}+\alpha_{k}}{\sum_{k=1}^{K}n_{m}^{k}+\alpha_{k}}  \\
\vec{\phi}_{k,s} &= \dfrac{n_{k}^{s}+\beta_{s}}{\sum_{s=1}^{S}n_{k}^{s}+\beta_{s}}  \\
\vec{\psi}_{k,o} &= \dfrac{n_{k}^{o}+\eta_{o}}{\sum_{o=1}^{O}n_{k}^{o}+\eta_{o}} \\
\vec{\varphi}_{k,t} &= \dfrac{n_{k}^{t}+\gamma_{t}}{\sum_{t=1}^{B}n_{k}^{t}+\gamma_{t}}.
\end{aligned}
\end{equation}

Based on the above analysis, the learning algorithm for TraLFM can be described in Algorithm 1. The major time consuming part in Gibbs Sampling for TraLFM is computing the conditional probability in Equation~(\ref{equ:newtopic}). Note that in TraLFM, we need to draw latent factor assignment for every occurrence of a trajectory unit in the trajectories, which costs time $O(K \ast M \ast \bar{T}_{s})$, where $M$ is the total number of trajectories in the training set, and $\bar{T}_{s}$ is the average length (in terms of the number of trajectory units) of a trajectory. We will evaluate the running time of TraLFM in the experiments.

\begin{algorithm}[h]
\small
\label{alg:gbs}
\caption{Learning Algorithm for \textit{TraLFM} }
\begin{algorithmic}[1]
\REQUIRE trajectory set $\cal T$, hyperparameters $\vec{\alpha}$, $\vec{\beta}$, $\vec{\eta}$, and $\vec{\gamma}$, number of latent factors $K$;
\ENSURE $\vec{\theta}$, $\vec{\phi}$, $\vec{\psi}$, $\vec{\varphi}$;
\FOR {each trajectory $T=m$}
\STATE assign a latent factor $k$ to each trajectory unit $tu$ randomly;
\STATE increase counts and sums of $n_{m}^{k}$, $n_{k}^{s}$, $n_{k}^{o}$, and $n_{k}^{t}$ respectively;
\ENDFOR
\FOR {each Gibbs Sampling iteration}
\FOR {each trajectory $T=m$}
\FOR {each $tu$}
\STATE decrease counts: $n_{m}^{k}-=1$, $n_{k}^{s}-=1$, $n_{k}^{o}-=1$, and $n_{k}^{t}-=1$;
\STATE decrease sums of $n_{m}^{k}$, $n_{k}^{s}$, $n_{k}^{o}$, and $n_{k}^{t}$;
\STATE draw a new latent factor according to Equation~(\ref{equ:newtopic});
\STATE increase counts and sums of $n_{m}^{k}$, $n_{k}^{s}$, $n_{k}^{o}$, and $n_{k}^{t}$ for the new assignment;
\ENDFOR
\ENDFOR
\STATE update $\vec{\theta}$, $\vec{\phi}$, $\vec{\psi}$, $\vec{\varphi}$ according to Equation~(\ref{equ:para2});
\ENDFOR
\STATE return $\vec{\theta}$, $\vec{\phi}$, $\vec{\psi}$, $\vec{\varphi}$;
\end{algorithmic}
\end{algorithm}

\section{Performance Evaluation}
In this section, we evaluate the performance of TraLFM with two real datasets. We first introduce the datasets, and then report the runtime of our model. Finally, we demonstrate the performances of TraLFM with two tasks: latent factor analysis and next location prediction.

\subsection{Data and Settings}
\textbf{VPR data}. We use a vehicle passage records dataset which is collected from the traffic surveillance system in a major metropolitan area (Jinan, China) with an area of 2,119 sq.km. Each record is extracted from the picture using optical character recognition (OCR), containing a vehicle ID (object: $o$), the location of the surveillance camera (location: $l$), and the time-stamp of object $o$ passing location $l$ (time: $t$). We take the records of 31 days, and pre-process them to form trajectories. To make the model more robust, we only consider trajectories that contain at least three locations. The detailed information is shown in Tab.~\ref{tab:data}. $\sharp$object means the number of unique objects, and the others have the similar meaning.

\textbf{Taxi data}. The taxi data is composed of all the complete trips of 442 taxis running in the city of Porto (Portugal) of 389 sq.km for a complete year (from 01/07/2013 to 30/06/2014)\footnote{http://www.geolink.pt/ecmlpkdd2015-challenge/dataset.html}. We discretize the region of interest into a grid with equal-sized cells, and assign a cell index for each GPS location. We process these data and describe the statistic result in Tab.~\ref{tab:data}.

\begin{figure*}[!thp]
\centering
\begin{minipage}{0.3\textwidth}
  \centering
  \includegraphics[width=.9\linewidth]{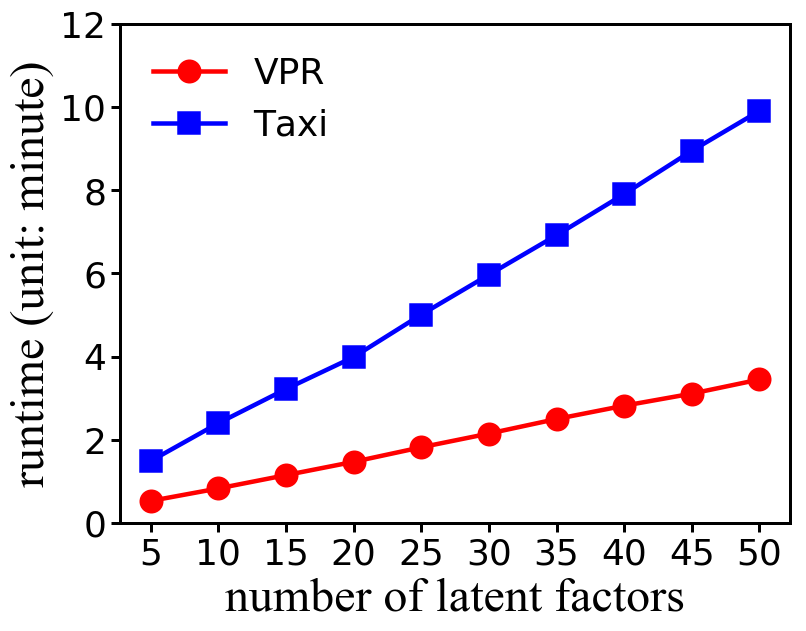}
  \captionof{figure}{Runtime of one hundred iterations}
  \label{fig:runtime}
\end{minipage}%
\hspace{0.03\textwidth}
\begin{minipage}{0.3\textwidth}
  \centering
  \includegraphics[width=.9\linewidth]{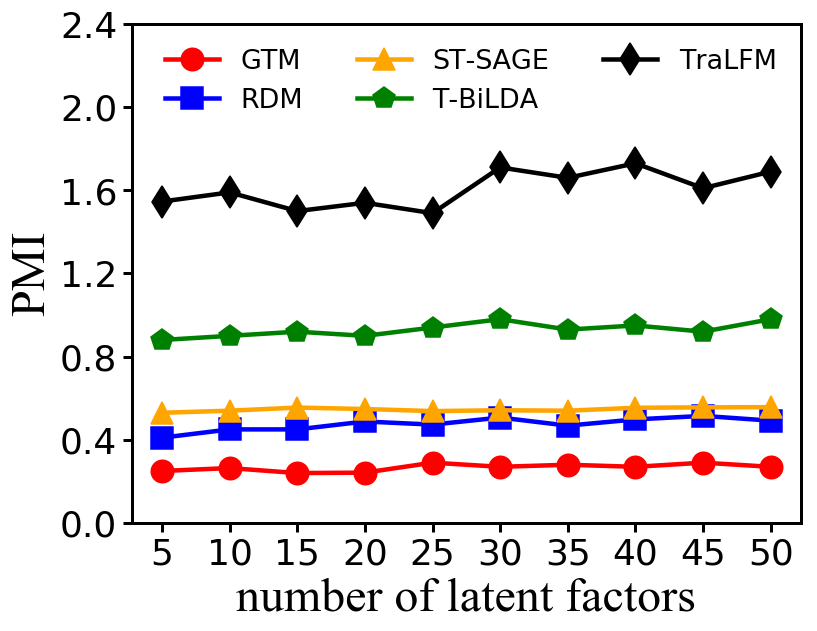}
  \captionof{figure}{Performance of latent factor coherence (VPR data)}
  \label{fig:pmivpr}
\end{minipage}%
\hspace{0.03\textwidth}
\begin{minipage}{0.3\textwidth}
  \centering
  \includegraphics[width=.9\linewidth]{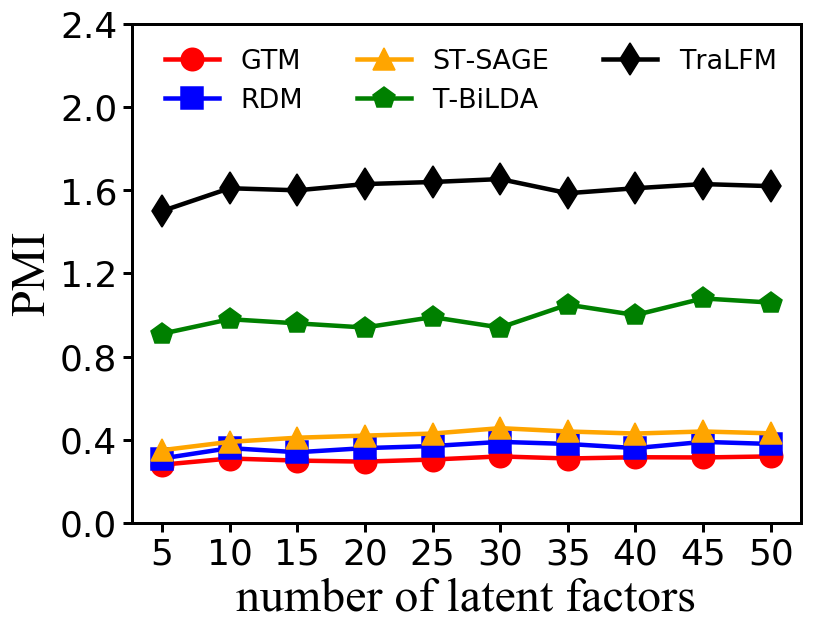}
  \captionof{figure}{Performance of latent factor coherence (Taxi data)}
  \label{fig:pmitaxi}
\end{minipage}%
\end{figure*}

\begin{table}
\centering
\renewcommand{\arraystretch}{1.2}
\caption{Data statistic.}
\begin{tabular}{ l| l | l}
\Xhline{1pt}
 & VPR data & Taxi data\\
\hline
$\sharp$objects &  35,005 &  442 \\
$\sharp$locations  & 225  &  13,907\\
$\sharp$trajectories &  927,175 & 164,609\\
$\sharp$first-order sequences &  8,619  &  163,252\\
$\sharp$second-order sequences &  106,871 & 572,050 \\
\Xhline{1pt}
\end{tabular}
\label{tab:data}
\end{table}

For all the experiments, we empirically set Dirichlet hyperparameters $\vec{\alpha}=\{50/K\}_{1}^{K}$, $\vec{\beta}=\{0.01\}_{1}^{S}$,  $\vec{\eta}=\{0.01\}_{1}^{O}$,  $\vec{\gamma}=\{0.01\}_{1}^{B}$ for simplicity, following the study of Griffiths and Steyvers \cite{griffiths2004finding}. We use 10-fold cross-validation and execute one hundred iterations of Gibbs sampling (TraLFM converges after about one hundred iterations in our experiments), and report the average results of different metrics to evaluate the performance. All the experiments are done on a 3.4GHz Intel Core i7 PC with 16GB main memory. The default values for the order of location sequences $r$, the number of latent factors $K$, and the size of time bin are 2, 40 and 2. We will evaluate the effect of these parameters in the experiments.

\subsection{Running Time}
Based on the learning algorithm for TraLFM, we know that the number of trajectory units in the training set and the number of latent factors determine the runtime of each iteration. Fig.~\ref{fig:runtime} shows the runtime of one hundred iterations for both datasets with different number of latent factors ($K$). On one hand, for the same $K$, the runtime with the Taxi dataset is larger than that with the VPR dataset, because the Taxi dataset has more trajectory units; on the other hand, the runtime on both datasets increases gradually with the rise of $K$. Note that, we could train the proposed TraLFM offline in advance, and use the learned distributions (e.g., $\vec{\phi}$, $\vec{\psi}$, $\vec{\varphi}$) in the real-time applications.

\subsection{Latent Factor Analysis}
As discussed in aforementioned sections, TraLFM provides an additional benefit of understanding human mobility behaviors. Therefore, we first examine the performance of TraLFM via analyzing the latent factors directly. Intuitively, we need to ask the objects (e.g., drivers in the VPR data) to understand their mobility patterns as the ground truth. However, it is unrealistic. So we choose an alternative method which is widely used in topic models to measure latent factor coherence based on the association of the sequences. Specifically, the latent factors learned from TraLFM are a multinomial distribution over sequences, and can be displayed by the $q$ most probable sequences therein. The top-$q$ sequences usually provide sufficient information to determine the interpretation of a latent factor, and distinguish one from another.

\subsubsection{Evaluation metrics}
We adopt the metric in topic models to measure the quality of the learned latent factors \cite{newman2011improving,tang2014understanding}.

\textbf{Point-wise Mutual Information} (PMI) is motivated by measuring the sequence association between all pairs of sequences in the top-$q$ sequences of each latent factor. PMI is defined as follows:
\begin{equation}
\small
\begin{aligned}
PMI(\vec{s}) &= \dfrac{2}{q(q-1)} \sum_{i<j} pmi(s_{i},s_{j}), ij\in \left\lbrace 1 \ldots q \right\rbrace   \\
pmi(s_{i},s_{j}) &= \log \dfrac{P\left( s_{i},s_{j} \right) }{P\left( s_{i} \right) P\left( s_{j} \right)},
\end{aligned}
\end{equation}
where $P\left( s_{i} \right)$ is the frequency of sequence $s_{i}$ occurring in all the trajectories and it can be computed in advance. We set $q$ at 10 according to the suggestion by Newman et al. \cite{newman2011improving}, and the number of $pmi(s_{i},s_{j})$ over the set of distinct sequence pairs in the top-10 sequences is 45. Apparently, a larger PMI indicates better latent factor coherence.

\subsubsection{Baselines}
We compare TraLFM with some state-of-the-art topic models including GTM \cite{long2012exploring}, RDM \cite{farrahi2011discovering}, ST-SAGE \cite{wang2017st} and T-BiLDA \cite{shen2009topic}.

\textbf{GTM} is a generative model based on LDA for discovering the geographic topics, in which the location represents a word and a user's trajectory represents a document.

\textbf{RDM} is a generative model for the automatic discovery of daily location-based routine patterns with LDA and the Author Topic model, in which it considers the object information.

\textbf{ST-SAGE} is a spatial-temporal sparse additive generative model, which takes into account both user interests and temporal dynamics of user behaviors.

\textbf{T-BiLDA} uses global transition probability and temporal information to refine the mixture distribution over topics for sequence analysis.

\begin{table*}[!th]
\small
\caption{Examples of discovered latent factors}
\renewcommand{\arraystretch}{1.2}
\centering
\begin{tabular}{l  l  l  l}
\Xhline{1pt}
 & \qquad \quad latent factor 1 & \qquad \quad latent factor 2 & \qquad  \quad latent factor 3 \\
\hline
                     &	5 \ [8:00-10:00@weekday]  &	9 \ [16:00-18:00@weekday]    &  20 [14:00-16:00@weekend]\\
                     &	4 \ [6:00-8:00@weekday]	 &  10 [18:00-20:00@weekday]	&  21 [16:00-18:00@weekend]\\
top-5 time bins      &  6 \ [10:00-12:00@weekday]	 &  11 [20:00-22:00@weekday]	&   11 [20:00-22:00@weekday]  \\
                     &  7 \  [12:00-14:00@weekday] &	8 \ [14:00-16:00@weekday]	&      19 [12:00-14:00@weekend]\\
                     &	17 [8:00-10:00@weekend]	 &  7 \ [12:00-14:00@weekday]   &    18 [10:00-12:00@weekend]\\
\Xhline{1pt}
\end{tabular}
\label{table:activity}
\end{table*}

\subsubsection{Performance of methods}
We compare TraLFM with the baselines (GTM, RDM, ST-SAGE, T-BiLDA) in both datasets,  and report the average PMI of each method in Fig.~\ref{fig:pmivpr} and Fig.~\ref{fig:pmitaxi}. We set the size of time bin at 2 hours, the order of sequence at 2, and vary the number of latent factors from 5 to 50. Note that only TraLFM makes use of the sequences, objects and time information jointly. As shown in Fig.~\ref{fig:pmivpr} and Fig.~\ref{fig:pmitaxi}, TraLFM outperforms the baselines significantly for each number of latent factors. GTM does not exploit personal and temporal information, so its performance is much worse than other methods. RDM considers the user interests, and ST-SAGE takes both personal and temporal factors into account, and they perform better than GTM. But they do not model the order of locations. T-BiLDA considers the location transition and mines sequences of temporal activities. We observe that it performs much better than GTM, RDM and ST-SAGE. Our TraLFM is capable of capturing the latent factors in terms of personal, sequential and temporal information that resides in the trajectories, and it performs the best.

Further, we list the top-5 time bins (containing the index and actual time) under each latent factor in Tab.~\ref{table:activity}. Latent factor 1 usually occurs from 6:00 to 12:00 in weekdays, and is very likely related to going to work; latent factor 2 mainly appears in the evening rush hours, likely corresponding to people coming off work; latent factor 3 often happens during the afternoon on weekends and the evening in weekdays, indicating people have similar mobility patterns (e.g., shopping and entertainment) in those time periods. We can see that the latent factors discovered by TraLFM demonstrate unique characteristics about time bins in an explainable manner.

\subsection{Next Location Prediction}
Understanding human mobility patterns can help make location prediction accurately, because people's decision about where to go next is dictated by these latent factors. Consequently, we choose to evaluate the performances of TraLFM further via the task of next location prediction.

Given a location sequence of an object, we could first learn his/her latent factors $\vec{\theta}$, and then predict the location that he/she will arrive at next. The probability of observing the next sequence $s_{n+1}$ by object $o$ with TraLFM is as follows:
\begin{equation}
\small
\label{equ:next}
p(s_{n+1}= s |T,o,t) \propto \sum_{k=1}^{K} \vec{\theta}_{k} \times \vec{\phi}_{k,s} \times \vec{\psi}_{k,o} \times \vec{\varphi}_{k,t}.
\end{equation}
Finally, we choose the location with the maximum probability as the predicted next location.

\subsubsection{Evaluation metrics}
To compare different models, we adopt a popular evaluation metric, namely, \textit{average precision}, to evaluate the performance \cite{chen2015mining}.


\textbf{Average precision} measures the predictive ability of a model. It is desirable to consider the order of the returned next locations, and it can be formulated with
\begin{equation}
\small
Average Precision = \dfrac{1}{W} \sum_{w=1}^{W} \dfrac{1}{w_{i}},
\end{equation}
where $w_{i}$ denotes the position of the actual next location in the predicted list for the $w$-th testing instance and $W$ is the total number of testing trajectories.

\subsubsection{Baselines}
In addition to these generative models (GTM \cite{long2012exploring}, RDM \cite{farrahi2011discovering}, ST-SAGE \cite{wang2017st} and T-BiLDA \cite{shen2009topic}), we also compare TraLFM with some discriminative methods including PrefixTP \cite{qiao2018predicting} and NLPMM \cite{chen2015mining} introduced in Section~\ref{relatedwork}, which are popular in the task of next location prediction.

\textbf{PrefixTP} mines frequent trajectory patterns of connected vehicles based on an efficient prefix-projection technique to predict next locations.

\textbf{NLPMM} builds upon two models: the Global Markov Model and the Personal Markov Model, and it considers both individual and collective patterns in prediction.

\begin{table}[!t]
\centering
\renewcommand{\arraystretch}{1.2}
\caption{Results of methods on VPR and Taxi data in terms of Average Precision.}
\begin{threeparttable}
\begin{tabular}{ p{0.08\textwidth}| p{0.07\textwidth}<{\centering}| p{0.07\textwidth}<{\centering}| p{0.07\textwidth}<{\centering}| p{0.07\textwidth}<{\centering} }
\Xhline{1pt}
\multirow{2}*{method} & \multicolumn{2}{c|}{VPR data} & \multicolumn{2}{c}{Taxi data}  \\
\cline{2-5}
\multirow{2}*{} & top-1 & top-5 &  top-1 & top-5 \\
\hline
GTM  & 0.046 & 0.102  & 0.007 & 0.011 \\

RDM  & 0.088 & 0.139   & 0.019  & 0.024 \\

ST-SAGE  & 0.096 & 0.152  & 0.023 & 0.028 \\

T-BiLDA  & 0.411 & 0.528  & 0.293 & 0.439 \\

PrefixTP & 0.422 & 0.536 & 0.365 & 0.529 \\

NLPMM  & 0.451 & 0.592 & 0.382 & 0.546 \\

\hline

TraLFM & \textbf{0.482}\tnote{1} & \textbf{0.606}\tnote{1}  & \textbf{0.399}\tnote{1} & \textbf{0.579}\tnote{1} \\

\Xhline{1pt}
\end{tabular}
\begin{tablenotes}
\footnotesize
\item[1] The improvements over the baselines are statistically significant in terms of paired t-test \cite{hull1993using} with $p$ value $<$ 0.01.
\end{tablenotes}
\end{threeparttable}
\label{tab:baseline}
\end{table}

\begin{figure*}[!t]
\centering
\subfigure[first-order sequences]{
\includegraphics[width=0.3\textwidth]{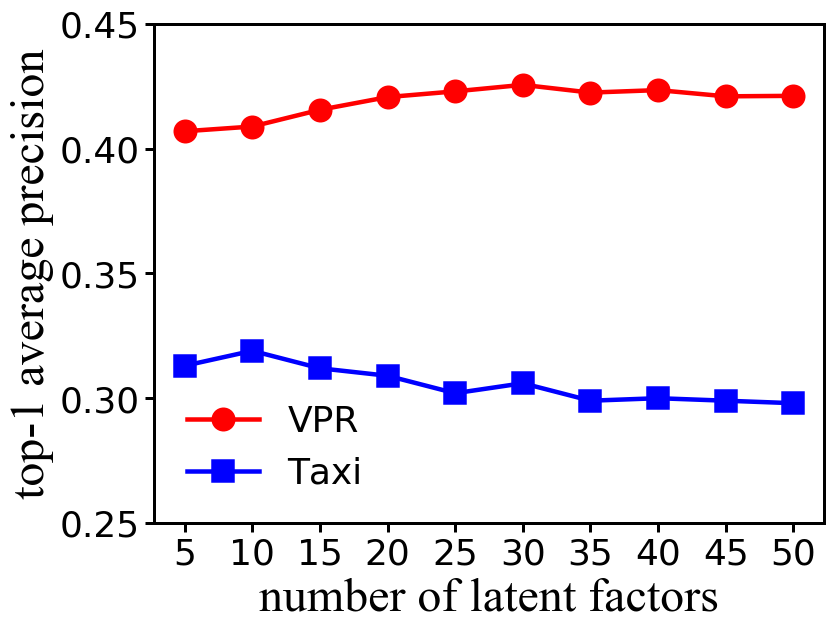}}
\subfigure[second-order sequences]{
\includegraphics[width=0.3\textwidth]{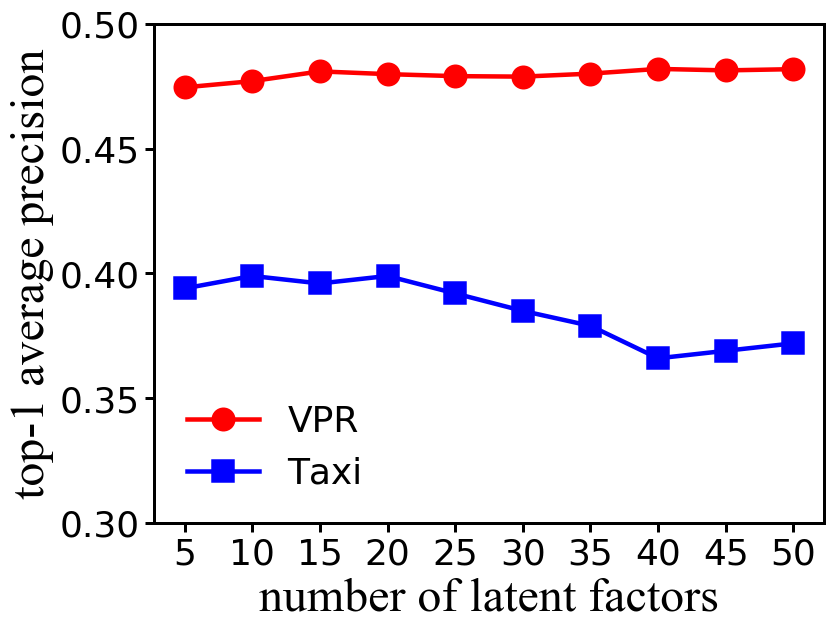}}
\subfigure[second-order sequences]{
\includegraphics[width=0.3\textwidth]{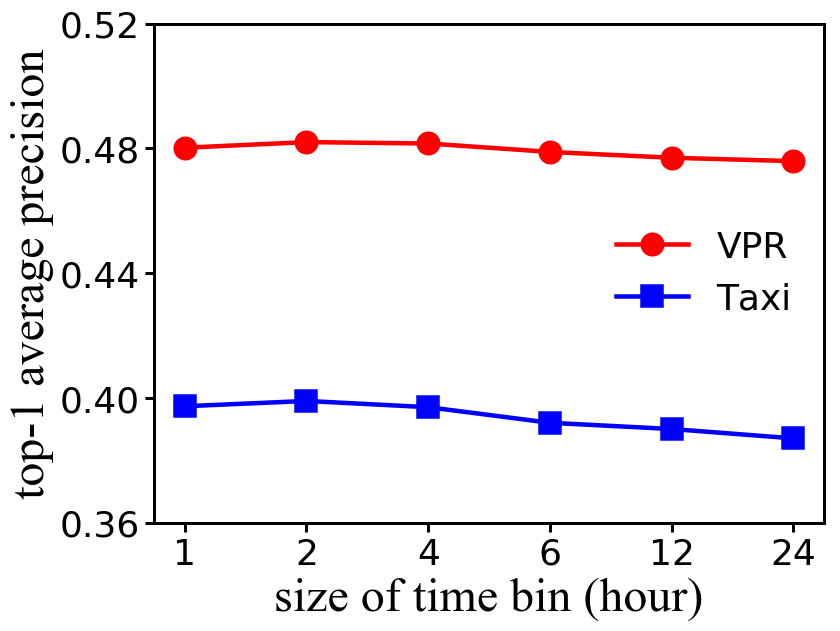}}
\subfigure[first-order sequences]{
\includegraphics[width=0.3\textwidth]{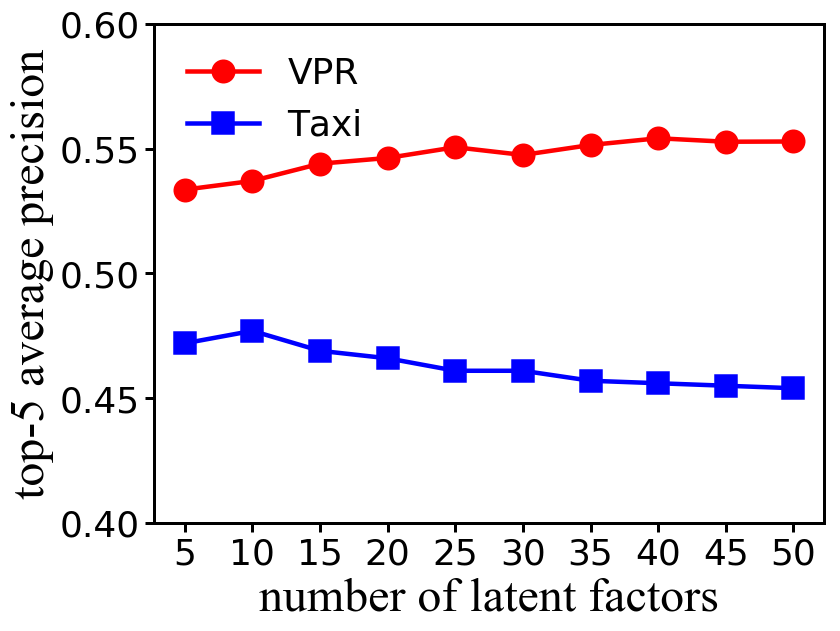}}
\subfigure[second-order sequences]{
\includegraphics[width=0.3\textwidth]{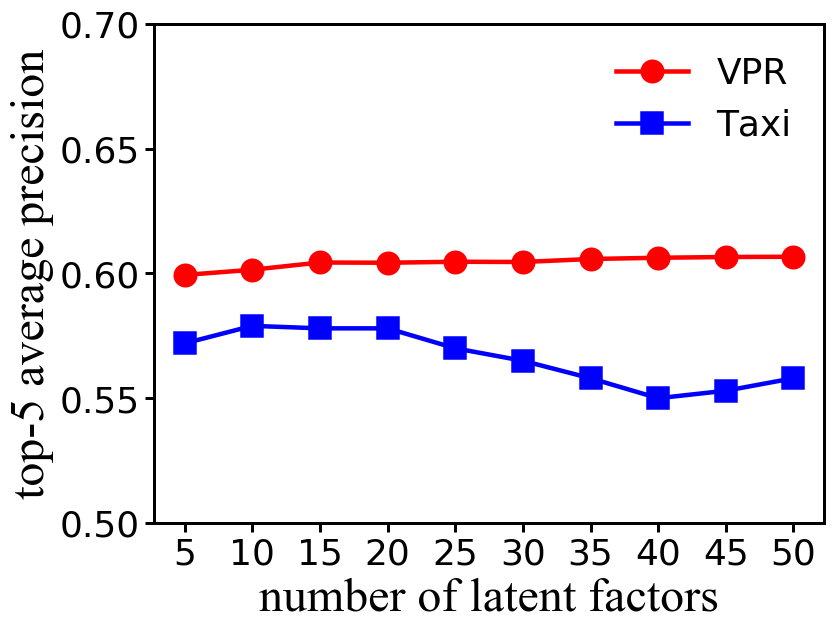}}
\subfigure[second-order sequences]{
\includegraphics[width=0.3\textwidth]{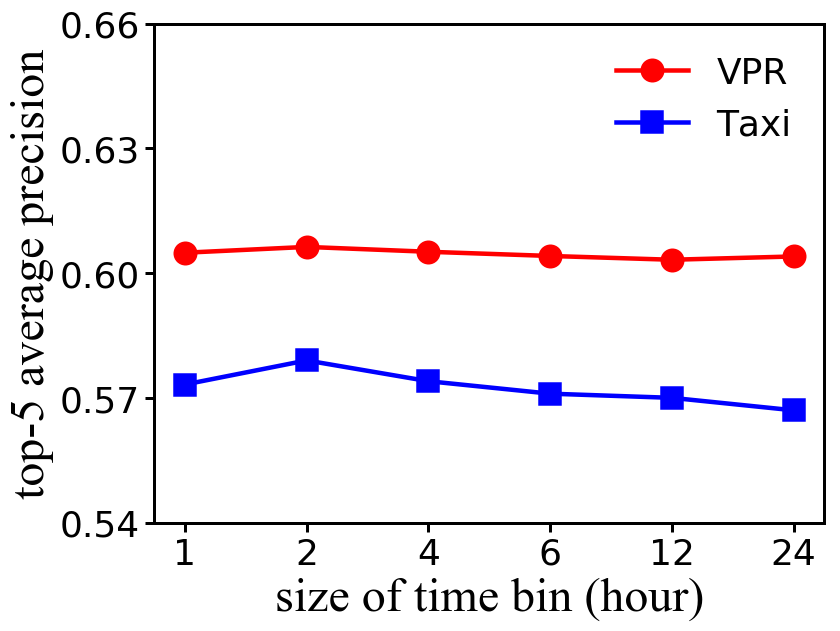}}
\caption{Empirical performance of tuning parameters. }
\label{fig:para}
\end{figure*}

\subsubsection{Performance of methods}

To evaluate the effectiveness of TraLFM further, we compare against the baseline methods with the task of predicting top-1 and top-5 next locations. We choose the optimal parameters for them after many experiments. The performance is shown in Tab.~\ref{tab:baseline}, and the best average precisions are highlighted in boldface. We notice that
\begin{enumerate}
\item All the methods perform better on the VPR data than on the Taxi data according to the average precision, as the routes taken by taxis are more diverse/random.
\item GTM, RDM and ST-SAGE have poor performance, as they only model the counts of unique locations in a trajectory and ignore the order of locations. Our proposed TraLFM considers the order of location sequences, and gains huge improvement. For example, compared with ST-SAGE, which has the best performance among the three methods, the top-1 and top-5 average precisions improve by 402.1\% and 298.7\% respectively on the VPR data, and by 1634.8\% and 1967.9\% on the Taxi data.
\item T-BiLDA significantly outperforms the above three methods, as it models the first-order sequences by using the global transition probability. But it does not model the objects, sequences and time simultaneously, so it performs worse than our TraLFM.
\item PrefixTP mines frequent trajectory patterns based on the previous trajectories of all the objects to predict successive locations, and NLPMM considers both individual and collective patterns. They both gain decent top-1 and top-5 average precisions. Compared with them, TraLFM can capture the latent factors based on the passed locations, and model the location sequences, objects and time jointly instead of treating them independently. Consequently, TraLFM performs the best, and the top-1 average precisions improve by 6.9\% on the VPR data and 4.5\% on the Taxi data compared with NLPMM.
\end{enumerate}

\subsubsection{Parameter Settings and Tuning}
We set three parameters in TraLFM, namely, the order of location sequences $r$, the number of latent factors $K$, and the size of time bin. Here we only show the performances of TraLFM with first-order and second-order sequences, as (1) the current location is only relevant to the immediately preceding locations, and (2) TraLFM may suffer from a few problems, e.g., serious data sparsity, higher time and space complexity, when processing higher order sequences. We then tune the parameters one by one on both datasets, and report the performances in Fig.~\ref{fig:para}.

We first set the size of time bin at 2 hours, and predict the next locations by varying the number of latent factors from 5 to 50. As shown in the Fig.~\ref{fig:para} (a), (b), (d), (e), 1) for both datasets, TraLFM with second-order sequences performs better than that with first-order, demonstrating the importance of the order of locations; 2) the top-1 and top-5 average precisions of TraLFM improve as we increase the number of latent factors and reach the peak values when $K= 40$ for VPR data and $K= 10$ for Taxi data. The reason for this performance degradation is that having too many or too few latent factors may hurt either the cohesiveness or the separation of the latent factors.

Next, we predict the next locations by varying the size of time bin (with second-order sequence, $K= 40$ for VPR data and $K= 10$ for Taxi data). We report the results in Fig.~\ref{fig:para} (c) and (f). As the size of time bin increases from 1 hour to 2 hours, the performance on both datasets improves. When we increase it further, the performance deteriorates. The reason is that a larger size of time bin makes the latent factors less time-specific.

\section{Conclusions}
In this paper, we have proposed a generative model called TraLFM to mine human mobility patterns via latent factor modeling from traffic trajectory data. TraLFM discovers the latent factors of a trajectory by jointly modeling location sequences, objects and time. The discovered latent factors help people understand human mobility patterns, potentially benefiting a dazzling array of applications, such as advertisement casting and next location prediction. We evaluate the performance of TraLFM via latent factor analysis and location prediction with two real datasets, and the experiments show that TraLFM is effective in understanding human mobility patterns and significantly outperforms the state-of-the-art methods in prediction tasks.

\bibliographystyle{IEEEtran}
\bibliography{wenxian}

\end{document}